\documentclass[submission, Phys]{SciPost}

\pdfoutput=1

\usepackage{ graphicx } 
\usepackage{ bm } 
\usepackage{ color }

\newcommand{\be}{\begin{equation}}
\newcommand{\ee}{\end{equation}}
\newcommand{\bea}{\begin{eqnarray}}
\newcommand{\eea}{\end{eqnarray}}

\begin{document}

\begin{center}{\Large \textbf{
Miraging a Majorana
}}\end{center}

\begin{center}
P. D. Sacramento\textsuperscript{1,2,*}
\end{center}

\begin{center}
{\bf 1} CeFEMA,
Instituto Superior T\'ecnico, Universidade de Lisboa, Av. Rovisco Pais, 1049-001 Lisboa, Portugal
\\
{\bf 2} Beijing Computational Science Research Center,
Beijing 100193, China
\\
* pdss@cefema.tecnico.ulisboa.pt 
\end{center}

\begin{center}
\today
\end{center}


\section*{Abstract}
{\bf
The image of a Majorana mode located on the focus of an elliptical corral
of free electrons is studied. The Majorana mode may be taken at the edge of
a topological wire superimposed on the two-dimensional electron gas.
At low energies the states of the wire are ignored except for the Majorana mode.
Usual tunneling to a fermionic mode is compared.  In the favorable
cases, tunneling to a Majorana mode leads to an enhanced mirage effect and a
spectral weight mainly confined around the foci, in comparison
to the tunneling to a fermionic mode. The Majorana character of the image is
shown computing the self-conjugacy.
}

\vspace{10pt}
\noindent\rule{\textwidth}{1pt}
\tableofcontents\thispagestyle{fancy}
\noindent\rule{\textwidth}{1pt}
\vspace{10pt}

\section{Introduction}
\label{sec:intro}

\begin{figure}[t]
\begin{center}
\includegraphics[width=0.55\textwidth]{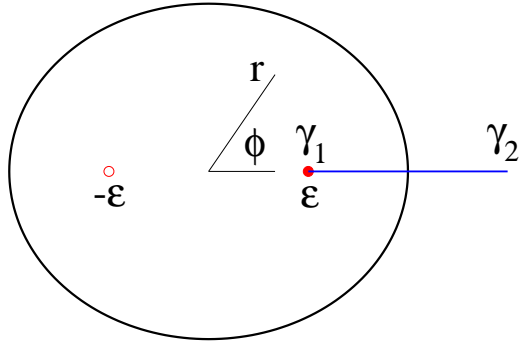}
\caption{\label{fig1}
(Color online)
Elliptical corral and wire with Majorana modes at the edges.
}
\end{center}
\end{figure}

Waves in confined geometries have complex interference patterns.
An interesting example is the enhanced image amplitude at one focus
of some confined geometry \cite{manoharan,fiete}, like elliptical
\cite{rossier,aligia0}, parabolic \cite{parabolic,parabolic2} or stadium corrals, of some state
with significant amplitude at the other focus. This phenomenon has been named
the mirage of a state. Examples in quantum mechanical systems have been found
both theoretically and experimentally such as the image of a Kondo resonance \cite{agam,bruno,morr2},
an impurity located at one of the foci \cite{kampf1,kampf2}, a magnetic impurity on a conventional
superconductor \cite{morr,aligia1,aligia2}.
Here we consider the corresponding situation of a Majorana mode placed at one focus
of an elliptical corral and study how it affects the electronic density of states of 
a system of free fermions and how the information of this localized state
is seen at the other focus. 

The problem of a Majorana fermion interacting with free electrons has attracted attention,
as a mean to its detection through the tunneling to and from the localized mode
\cite{bolech,liu,fidkowski}, since it has been shown that there is non-vanishing tunneling
\cite{semenoff}. 
Another interesting aspect is the possibility to show, detect and use
its intrinsic non-locality \cite{tewari,nilsson1,flensberg,stepanyuk}, since Majoranas
appear in pairs and form highly non-local regular fermions and, therefore, may be
used in long-distance teleportation \cite{fu1,fu2,fu3}. The study of interference
patterns on the surface of a topological insulator has also been considered recently \cite{franz},
as have the mirages created by Andreev reflected states \cite{su}.
Also, the non-abelian character and robustness due to topological protection of Majorana
modes, seems to offer a potentially useful way to implement topological quantum
computation \cite{kitaev,nayak}.  
The braiding of the Majorana modes and, in general, their manipulation, has attracted
considerable attention \cite{aliceab,hassler,necklace}.
One example has been the manipulation of Majorana modes
in structures involving collections of topological wires, for instance
applying external fields that change the topological nature \cite{bela},
such as pairing or chemical potential profiles, for instance using
shuttles that carry Majorana modes as some domain wall is displaced along
some wire \cite{franz2}.
The manipulation of the modes typically involves decoherence,
such as due to quasiparticle poisoning, which is particularly
serious if the manipulation of the Majorana modes is not carried out slow enough,
and requires some error correction, also due to some thermal noise
\cite{loss,noise,mcarlo}.
However, the manipulation of the Majorana modes has been proposed to provide a way
to establish networks for quantum communication \cite{lang} at arbitrary distances.
The possible existence of a mirage of a Majorana mode may lead to the possibility 
of manipulation of the Majorana modes without transport,
and may offer a way to avoid or decrease the decoherence, with a transmission of information
through the states in the corral. In this work we consider the conditions for its
appearance.

The Majorana mode may arise in different situations such as
at the ends of some topological wire. Examples are semiconductor wires with spin-orbit
coupling proximity coupled to a conventional superconductor and in the presence
of some Zeeman term \cite{kane,lutchyn,oreg}, or chains of magnetic impurties on top of a conventional or
triplet superconductor, either forming some magnetic spiral order or ferromagnetically
aligned with the extra effect of spin-orbit coupling, or even other orderings 
\cite{perge,perge2,glazman,klinovaja,us}.
One possible system to study would be to consider a two-dimensional conventional
superconductor with spin-orbit coupling, placing a ferromagnetically ordered 
long chain of magnetic impurities on
top of the superconductor, such that one of its ends is placed on one of the foci of
a corral. The corral may be implemented as in the study of a single impurity at the focus \cite{morr} 
placing a series of potential scattering impurities along an ellipse.
Since the magnetic chain has to be
sufficiently long so that the edge states do not overlap significantly and the edge modes
have vanishing energies, the system to be studied is large and numerically expensive.
It is therefore easier to analyse a situation in the continuum as represented in Fig. {\ref{fig1}.
We consider a system of electrons with a parabolic band in the confined geometry of an elliptical corral, implemented
imposing open boundary conditions, in interaction with a wire that is assumed to support some
Majorana mode at its edges. One of the edges is outside the corral and assumed far enough and,
since we are interested in the low-energy properties, we assume the wire can be simulated
exclusively by the Majorana mode, $\gamma_1$, located at the edge inside the corral.

The problem of free fermions can be solved exactly. One way is to use Mathieu
functions \cite{rossier}. Here we adopt, for simplicity, a description in terms of cylindrical Bessel functions.
These are the solutions for a circular corral, in which case the
two foci coincide. Introducing the eccentricity of the ellipse the Bessel functions are no longer
eigenfunctions, but one can construct an Hamiltonian matrix which is tridiagonal, if we
use properly rescaled variables \cite{nakamura}. Specifically, let us define the cartesian coordinates
in a two-dimensional system as $x/a= r \cos (\varphi)$, $y/b= r \sin(\varphi)$. Here $a$ and $b$
are the major and minor axis of the ellipse, $r$ is a radial coordinate and $\varphi$ the angular
coordinate. Given the eccentricity of the ellipse, $\epsilon$, the distance of the foci to the
origin is $f=a \epsilon$, and $\sigma=b/a=\sqrt{1-\epsilon^2}$. We will consider the case of $\epsilon=1/2$.
We solve Laplace's equation using the basis of the circular problem \cite{nakamura,aligia0}.
We consider the basis of normalized states 
$\Phi_{k,n}=\langle \vec{r}|kn\rangle = R_{kn} J_k(\gamma_{kn} r) e^{ik\varphi}$.
The function $J_k$ is the cylindrical Bessel function of order $k$, $R_{kn}$ ensure
that the basis is orthonormalized and $\gamma_{kn}$ is the $n$ zero of the Bessel function
of order $k$. 
Details of the method are presented in the appendix.
The solution of the free electron problem in the elliptical geometry
allows to identify states that have significant amplitudes at both foci,
as required to obtain a significant mirage effect.

\begin{figure}
\begin{center}
\includegraphics[width=0.49\textwidth]{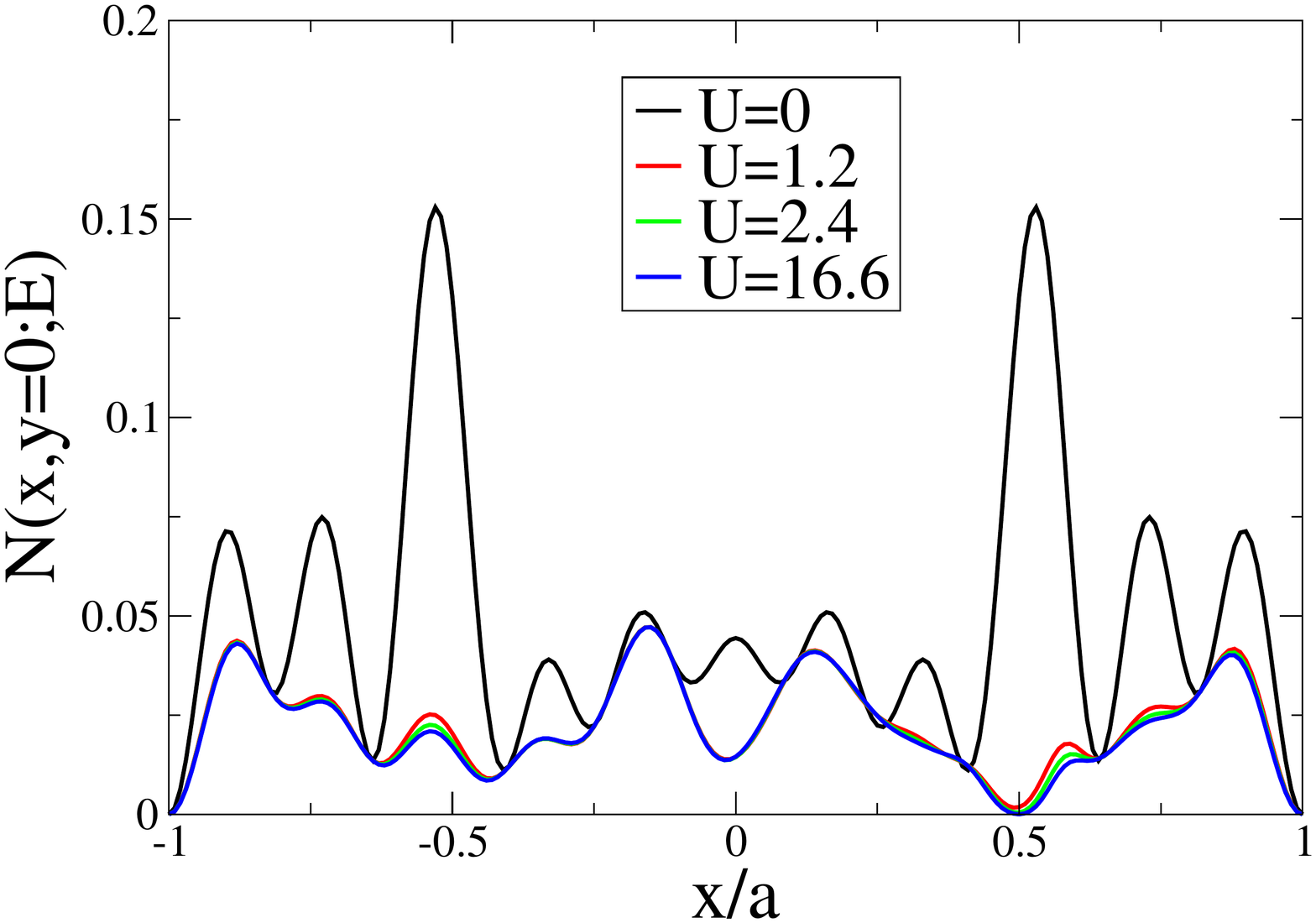}
\includegraphics[width=0.49\textwidth]{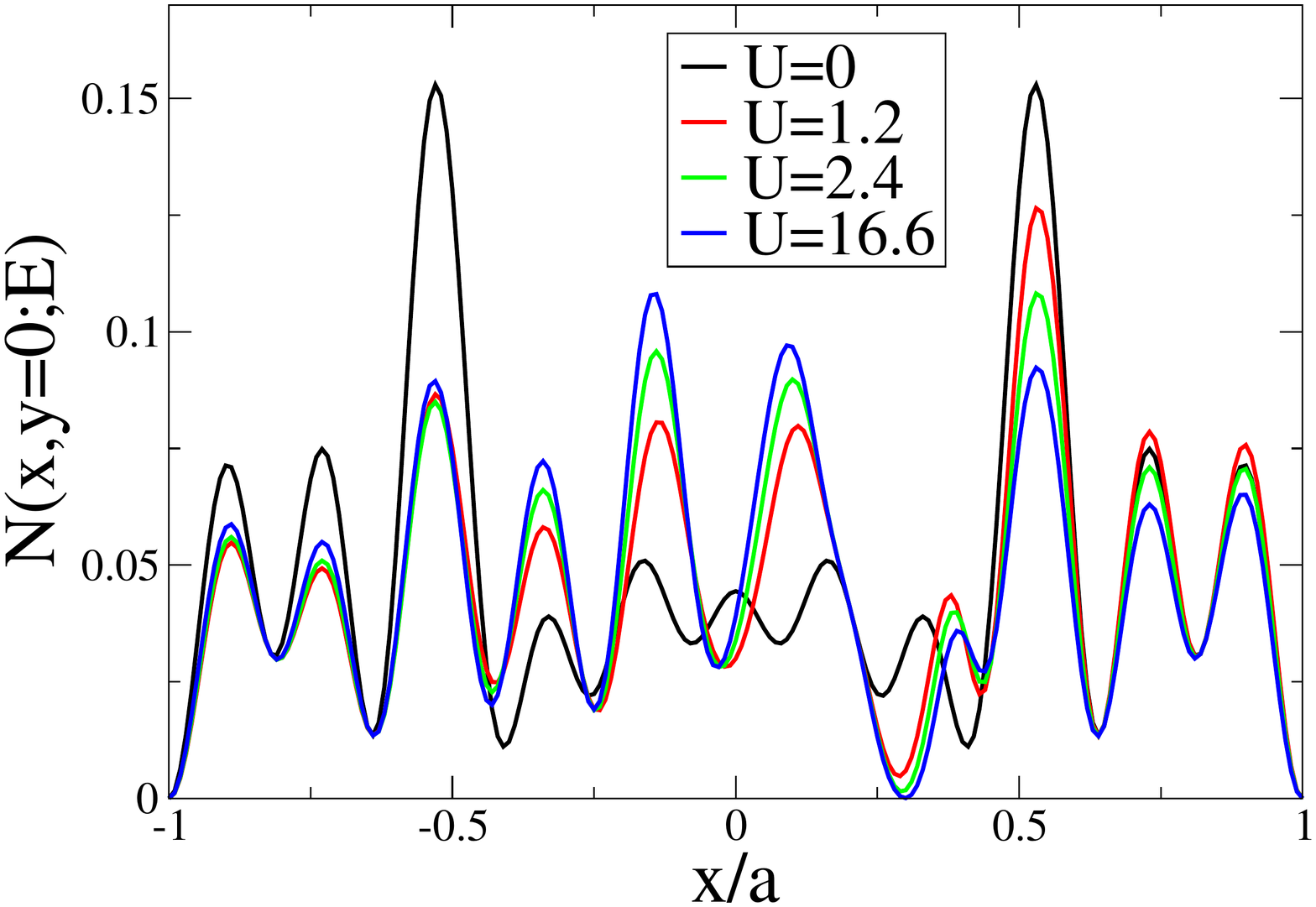}
\caption{\label{fig2}
(Color online)
Local density of states of a corral 
with a potential impurity, $U$, located at
a) right focus and b) off focus. The energy $E$ is chosen such that the
density of states is high at both foci when the impurity is absent, $U=0$.
}
\end{center}
\end{figure}

In section 2 the case of a simple potential impurity is briefly reviewed and the
local density of states is calculated. In section 3 the tunneling to a fermionic
or a Majorana mode at the edge of a wire superimposed on the elliptical corral is considered.
The local density of states (LDOS), differential LDOS, and the self-conjugacy of the lowest energy modes are
calculated, showing the enhanced mirage effect and Majorana character of the image at
the other focus. In section 4 the case of two wires superimposed on the corral, with one edge
at one or the other focus, shows the Majorana character at both wires, provided there is a
Majorana mode in one of them. In section 5 the conclusions are presented.

\section{Local impurity in elliptical corral}

Let us consider first a two-dimensional system of free electrons in the presence
of the corral without the wire, but with one impurity added \cite{kampf1} at site
$\vec{r}_i$ described by an Hamiltonian
\be
H = \int d^2 r c^{\dagger}(\vec{r}) \left(-\frac{\hbar^2}{2m} \nabla^2 \right) c(\vec{r})
+U \int d^2 r \delta(\vec{r}-\vec{r}_i) c^{\dagger}(\vec{r}) c(\vec{r})
\ee

Diagonalizing the Hamiltonian in the basis of Bessel functions,
one may calculate the local density of states in the corral.
The LDOS at point $\vec{r}$ and energy $E$ may be obtained as
\be
N(\vec{r},E) = \frac{1}{\pi} \sum_j \psi_j(\vec{r}) \psi_j^*(\vec{r}) \frac{\delta}{(E-E_j)^2+
\delta^2}
\ee
where $\psi_j(\vec{r})$ are the eigenstates, $E_j$ the energy eigenvalues and $\delta$ is of the order of
the lowest energy of the free problem, $U=0$. The energy $E$ is selected at a given $E_j$ such that
its wave function has a large amplitude at the foci locations \cite{evalue}. 

In Fig. \ref{fig2} we place
the impurity at the focus $\vec{r}_i=(r=1/2,\varphi=0)$ or the impurity
off focus at $\vec{r}_i=(r=0.3,\varphi=0)$. The effect of the repulsive potential at the impurity location is to
reduce the LDOS. This is seen both when the impurity is placed on focus or off focus.
The effect is clearly enhanced as the impurity strength, $U$, increases, but
even a moderate impurity potential significantly reduces the density of states
at the right focus. Both figures are obtained
for the density of states calculated at an energy, $E$, that is such that, when the impurity is absent,
there is a high spectral weight at both foci of the ellipse. As shown before, this is essential
to have a strong mirage effect. 
In Fig. \ref{fig2} the mirage effect may be understood by the decrease of
the density of states at $x=-\epsilon, y=0$ due to the potential. 
Note that the mirage is not perfect even though the impurity
potential is large. Placing the impurity off focus at $r=0.3$, 
no such decrease of the density of states is observed at $x=-0.3,y=0$. 
Indeed there is an anti-mirage
effect \cite{kampf1}, in the sense that the density of states is larger than that in the absence of the impurity.
This different behavior highlights the significance of the mirage effect of the focus location.

\section{Hybridization to wire edge: fermionic vs. Majorana modes}

\begin{figure}
\begin{center}
\includegraphics[width=0.4\textwidth]{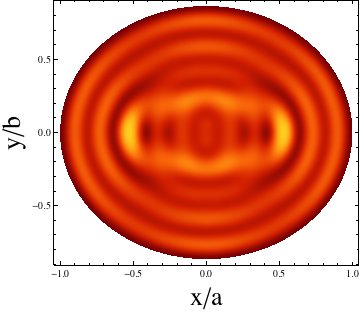}
\includegraphics[width=0.52\textwidth]{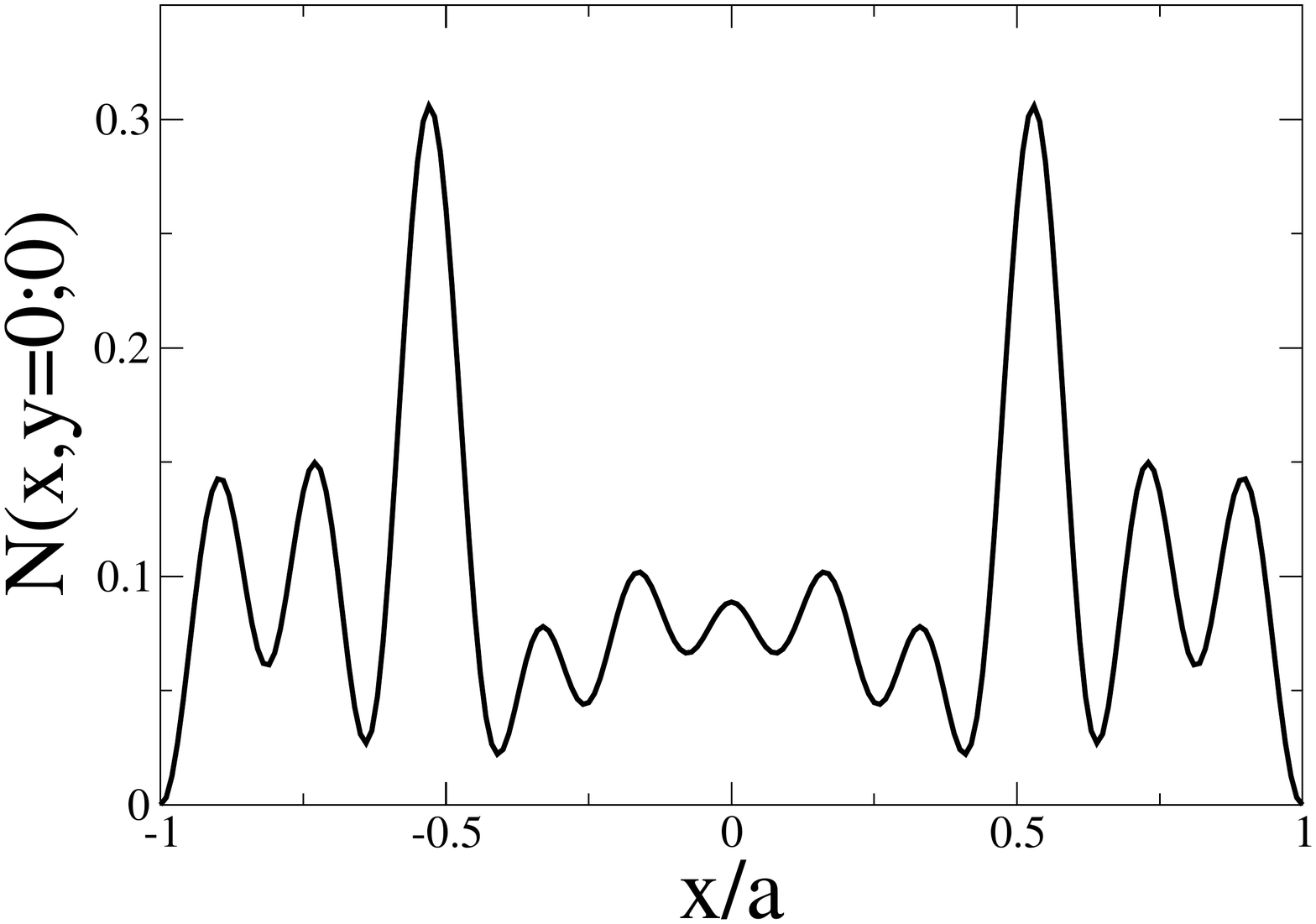}
\caption{\label{fig3}
(Color online)
Conduction electron contribution to a) two-dimensional local density of states with a chemical potential chosen at the energy, $E$, of Fig. \ref{fig2} 
with the wire uncoupled, $t=0,t_M=0$.
The lighter color corresponds to the larger
density of states. 
b) A cut along the major axis of the ellipse. 
}
\end{center}
\end{figure}

Let us now consider the situation with the wire in contact with the two-dimensional
fermionic gas, with its edge located at $\vec{r}_e$ and with no impurity present. 
The two edge Majorana modes couple to form a regular fermion
as $d=1/2(\gamma_1 + i \gamma_2)$. We consider now an Hamiltonian of the form
\bea
H &=& \int d^2 r c^{\dagger}(\vec{r}) \left(-\frac{\hbar^2}{2m} \nabla^2 -\mu \right) c(\vec{r})
\nonumber \\
&+& \int d^2 r \delta (\vec{r}-\vec{r}_e) t \left(d+d^{\dagger} \right) \left(
c(\vec{r})-c^{\dagger}(\vec{r}) \right)
\nonumber \\
&+& \xi_R \left( 2 d^{\dagger} d -1 \right)
\label{hmatrix}
\eea
The term that couples the electrons in the two-dimensional system with the Majorana mode
at the edge of the wire may be expanded as
\be
t \left(d+d^{\dagger} \right) \left(
c(\vec{r})-c^{\dagger}(\vec{r}) \right) 
\rightarrow t \left( c^{\dagger} d + d^{\dagger} c \right) +
t_M \left(d c +c^{\dagger} d^{\dagger} \right)
\ee
The first term is the usual hopping to the edge of the wire. In the case of a Majorana
$t=t_M$. Varying $t_M$ we may tune the mode from a Majorana to a usual tunneling between
regular fermionic states, obtained when $t_M=0$.
$\xi_R$ is the exponentially small coupling between the two edges of the wire
and $\mu$ is the chemical potential.

\begin{figure}
\begin{center}
\includegraphics[width=0.37\textwidth]{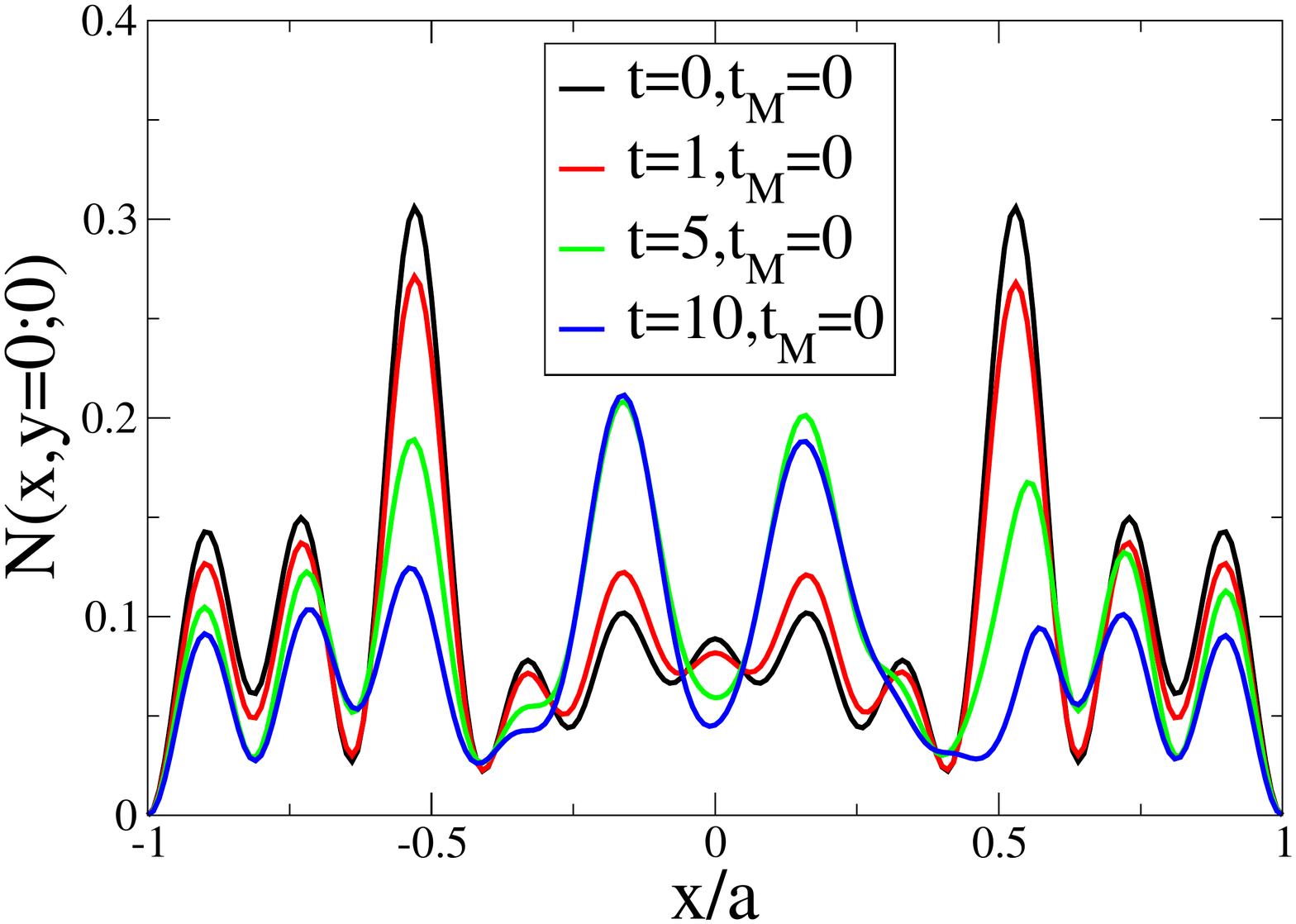}
\includegraphics[width=0.37\textwidth]{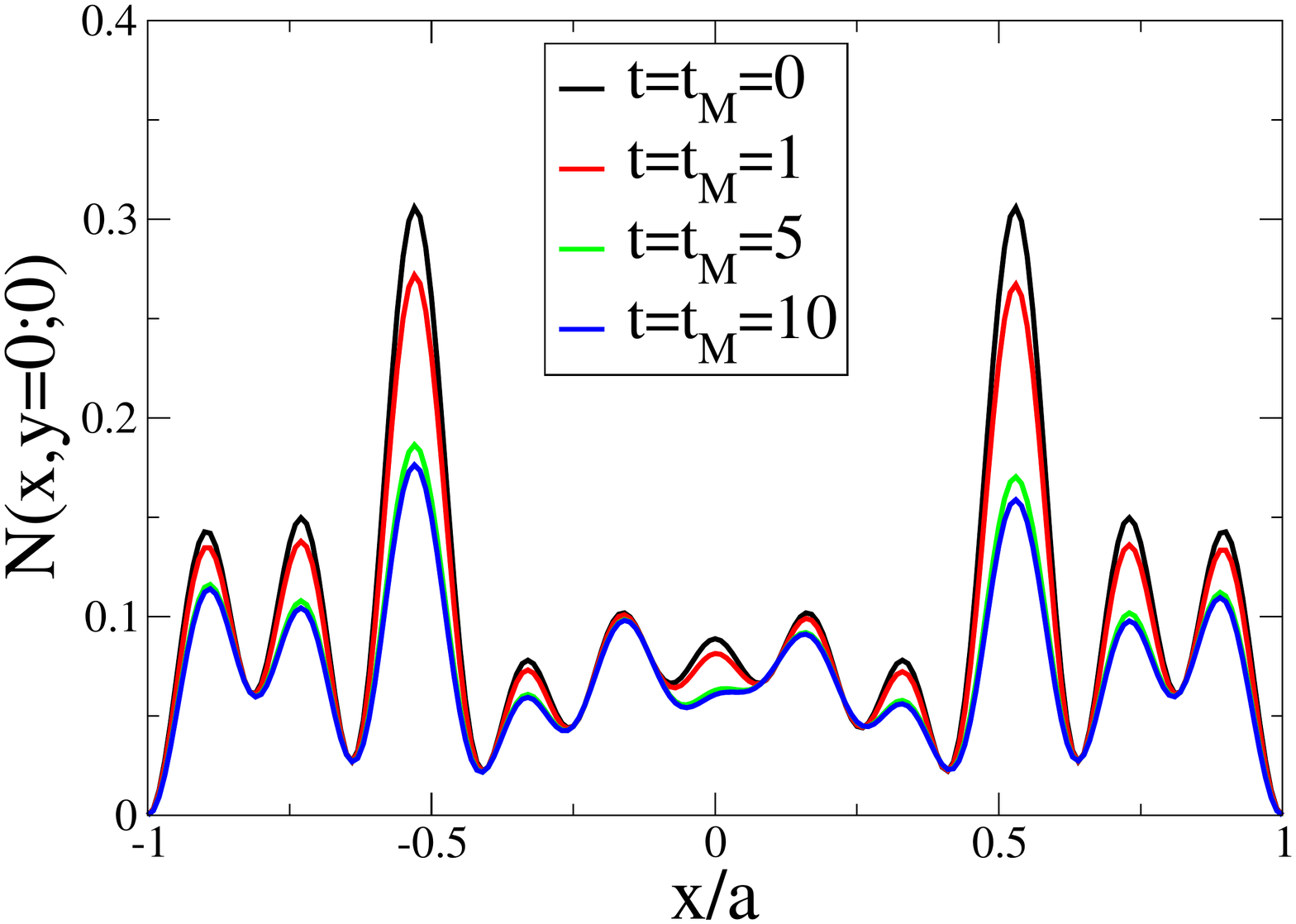}
\includegraphics[width=0.3\textwidth]{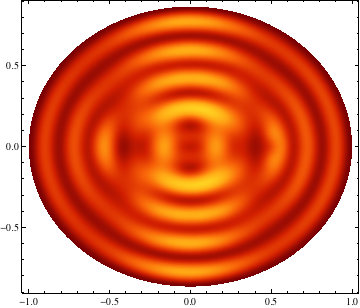}
\includegraphics[width=0.3\textwidth]{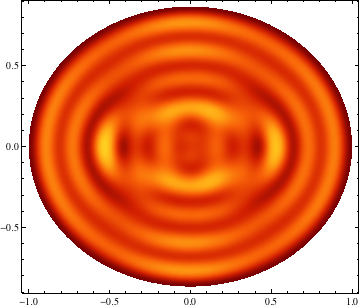}
\caption{\label{fig4}
(Color online)
Conduction electron contribution to the local density of states for a cut along the major axis in the case that the edge of the wire 
is placed at the right focus for
a) a fermionic mode at the edge $t_M=0$ and $t=0,1,5,10$ and b) a Majorana mode at
the edge $t=t_M=0,1,5,10$.  
Two-dimensional local density of states with the edge of the wire placed at the right focus for
c) a fermionic mode at the edge $t=5,t_M=0$ and d) a Majorana mode at
the edge $t=t_M=5$. The axis are as in Fig. 3.
}
\end{center}
\end{figure}

We may diagonalize the problem and determine the wave functions.
These may be calculated expanding the wave functions of the electrons
on the ellipse as
\bea
u(r,\varphi) &=& \sum_{k n} u_{kn} \Phi_{kn}(r,\varphi)
\nonumber \\
v(r,\varphi) &=& \sum_{k n} v_{kn} \Phi_{kn}(r,\varphi)
\eea
In addition there is a state localized at the edge of the wire, characterized by the wave functions
$u_d$ and $v_d$.

Introducing the vector $\psi= \left(u_{kn}, u_d, v_{kn}, v_d\right)^T$, 
we may write
the Hamiltonian matrix as
\be
\left(\begin{array}{cccc}
 (H_0)_{k^\prime,n^{\prime};k,n} &  t \Phi_{k^{\prime},n^{\prime}}^e &   0 &  t_M \Phi_{k^{\prime},n^{\prime}}^e \\ 
 t \Phi_{k,n}^e &  2 \xi_R &  -t_M \Phi_{k,n}^e &  0 \\ 
 0 &  -t_M \Phi_{k^{\prime},n^{\prime}}^e &  -(H_0)_{k^\prime,n^{\prime};k,n} &  -t \Phi_{k^{\prime},n^{\prime}}^e \\ 
 t_M \Phi_{k,n}^e &  0 &  -t \Phi_{k,n}^e &  -2\xi_R \\ 
 \end{array}\right) 
\label{hmatrix}
\ee
Here $(H_0)_{k^\prime,n^{\prime};k,n}$ is the matrix element in the Bessel function
basis of $-\hbar^2/(2m) \nabla^2-\mu$. Also, $\Phi_{k,n}^e$ is the basis function at
the edge location, $\vec{r}=\vec{r}_e$, which may coincide with one focus or can be placed off-focus.
We use energy units such that $\hbar^2/\left(2ma^2\sigma^2 \right) =1$. In these units
the lowest energy is of the order of $\gamma_{01}^2$, square of the first zero of the Bessel function of
order $0$.

\subsection{Local density of states}

Having solved the problem we may now calculate the LDOS in the corral.
This can be obtained using
\bea
N(\vec{r},\omega) &=& 
\frac{1}{\pi} \sum_j 
\frac{\delta}{(\omega-E_j)^2+ \delta^2} 
\left( |u_j(r,\varphi)|^2 + \delta(\vec{r}-\vec{r}_e) |u_d|^2 \right)
\nonumber \\
&+& \frac{1}{\pi} \sum_j 
\frac{\delta}{(\omega+E_j)^2+ \delta^2} 
\left( |v_j(r,\varphi)|^2 + \delta(\vec{r}-\vec{r}_e) |v_d|^2 \right)
\eea
The energies, $\omega, E_j$, are now measured with respect to the chemical potential.
There are two contributions to the LDOS one coming from the conduction electrons
in the corral and the other from the localized state at the edge location.
In most results shown ahead we will present results for the LDOS of the
conduction electrons.

In Fig. \ref{fig3} we present results for conduction electron contribution to the LDOS
with the chemical potential
taken as $\mu=E$, for a cut along $x$.
This corresponds to the case of Fig. \ref{fig2} with no impurity, $U=0$.

\begin{figure}
\begin{center}
\includegraphics[width=0.24\textwidth]{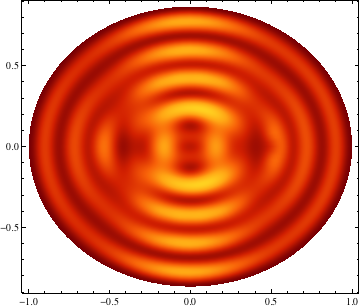}
\includegraphics[width=0.24\textwidth]{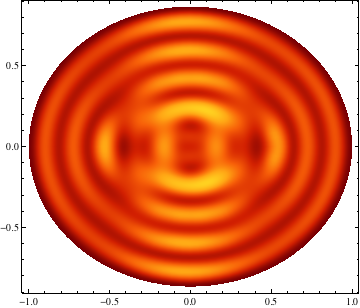}
\includegraphics[width=0.24\textwidth]{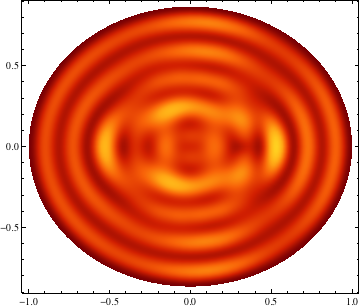}
\includegraphics[width=0.24\textwidth]{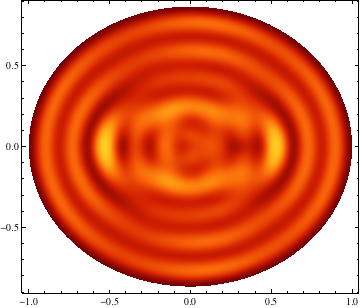}
\caption{\label{fig5}
(Color online)
In the two left panels the local density of states is shown taking
$\mu$ between two energy levels, for $t=5,t_M=0$
and $t=t_M=5$, respectively. In the two right panels $\mu$ is located at
the energy level, $E$, but the wire edge is now placed off-focus. The same tunneling amplitudes are considered, 
respectively.  The axis are as in Fig. 3.
}
\end{center}
\end{figure}

Let us now include the coupling between the wire and the electrons in the corral.
In Fig. \ref{fig4} we consider the edge of the wire at the right focus and compare
the tunneling to a fermionic mode with the tunneling to a Majorana mode, as a function of the tunneling
amplitudes. At the foci the effect is to reduce the density of states, in increasing
rate as the tunneling amplitude increases. In the case of the tunneling to a fermionic mode
the LDOS increases in the region near the center of the ellipse while
tunneling to a Majorana mode this amplitude decreases with respect to the uncoupled wire.
The character of the state of a large amplitude at the foci
and small in between is preserved in the case of the tunneling to a Majorana.
This behavior is clearly shown in Fig. \ref{fig4} where the full two-dimensional local
density of states is shown: the pattern of the tunneling to a Majorana mode displays the
peaks at both foci while the tunneling to a fermionic mode shows a spreading of the density of states
over the corral and the mirage effect is lost.

The results shown so far were obtained fixing the chemical potential at a given eigenstate of the
system with no wire and no impurity, specifically the level considered in Fig. 2 and placing the
edge of the wire on the right focus.
In Fig. \ref{fig5} we calculate the LDOS 
taking the chemical potential between the state with energy $E$, considered above, and the next
energy level of the free system. In the case of the tunneling to a fermionic
mode, there is a decreased density of states at the foci and a considerable distribution
along the smaller axis of the ellipse. On the other hand, in the case of tunneling to
the Majorana mode, while there is still a large density of states at the foci, there is also
a large density of states along the smaller axis and the character of the peaks centered
at the foci is somewhat lost. In the same figure we also consider the density of states
when the edge of the wire is off-focus and pinning again the chemical potential at the energy
level of Fig. \ref{fig2}. We find that in the case of tunneling to a fermionic mode or a
Majorana mode, the results are somewhat similar and the density of states does not change
appreciably from the uncoupled wire case. A similar result is found if we pin the chemical potential
to some energy level of the free system such that its wavefunction is not significantly
peaked at the foci. In these cases the influence of the extra state at the edge of the wire is
not very strong and the wavefunctions/density of states are not appreciably changed. This occurs also
if the wire is placed off-focus. Therefore the mirage effect requires some fine tuning, as expected.

Taking into account the coupling between the two Majorana modes at the two edges of
the wire ($\gamma_1$ and $\gamma_2$ in Fig. \ref{fig1}), which implies taking $\xi_R \neq 0$
does not change qualitatively the results, if the chemical potential is pinned at the same energy level.
However, shifting the chemical potential between this level and the next, has the effect that
the mirage effect is lost and the density of states is similar to the coupling to a fermionic
mode. If the chemical potential is pinned between two free energy levels, even though there
is a coupling to a Majorana-like mode (in the sense that $t=t_M$) the Majorana
character is lost, since the two Majoranas of the wire are coupled and there is a gap
in the spectrum. However, even if $\xi_R \neq 0$ but the chemical potential is pinned
on one free energy level, there is a zero energy mode in the spectrum.
However, this zero energy mode is not a true Majorana since the lowest state is
not self-conjugate.

\begin{figure}
\begin{center}
\includegraphics[width=0.38\textwidth]{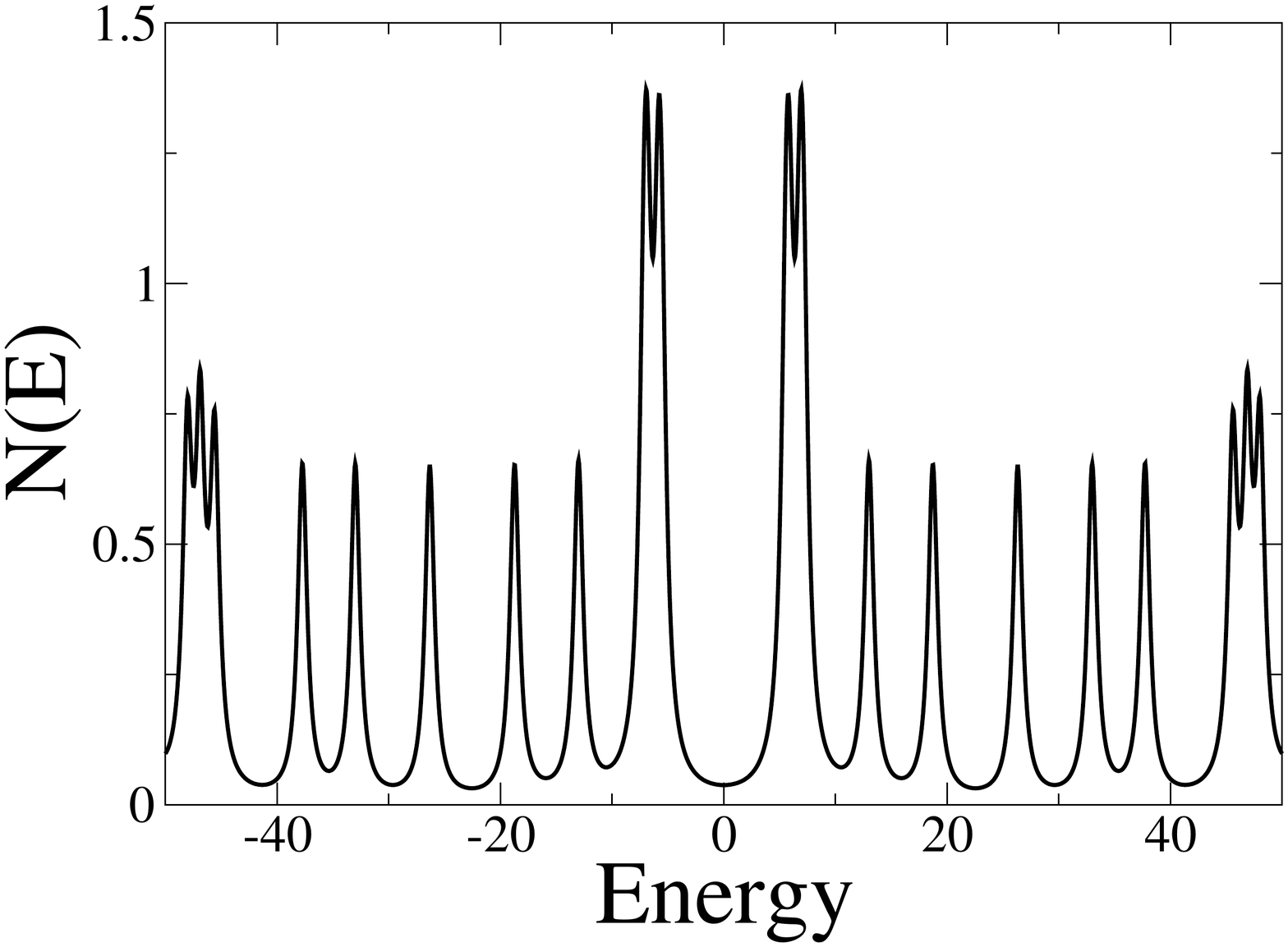}
\includegraphics[width=0.38\textwidth]{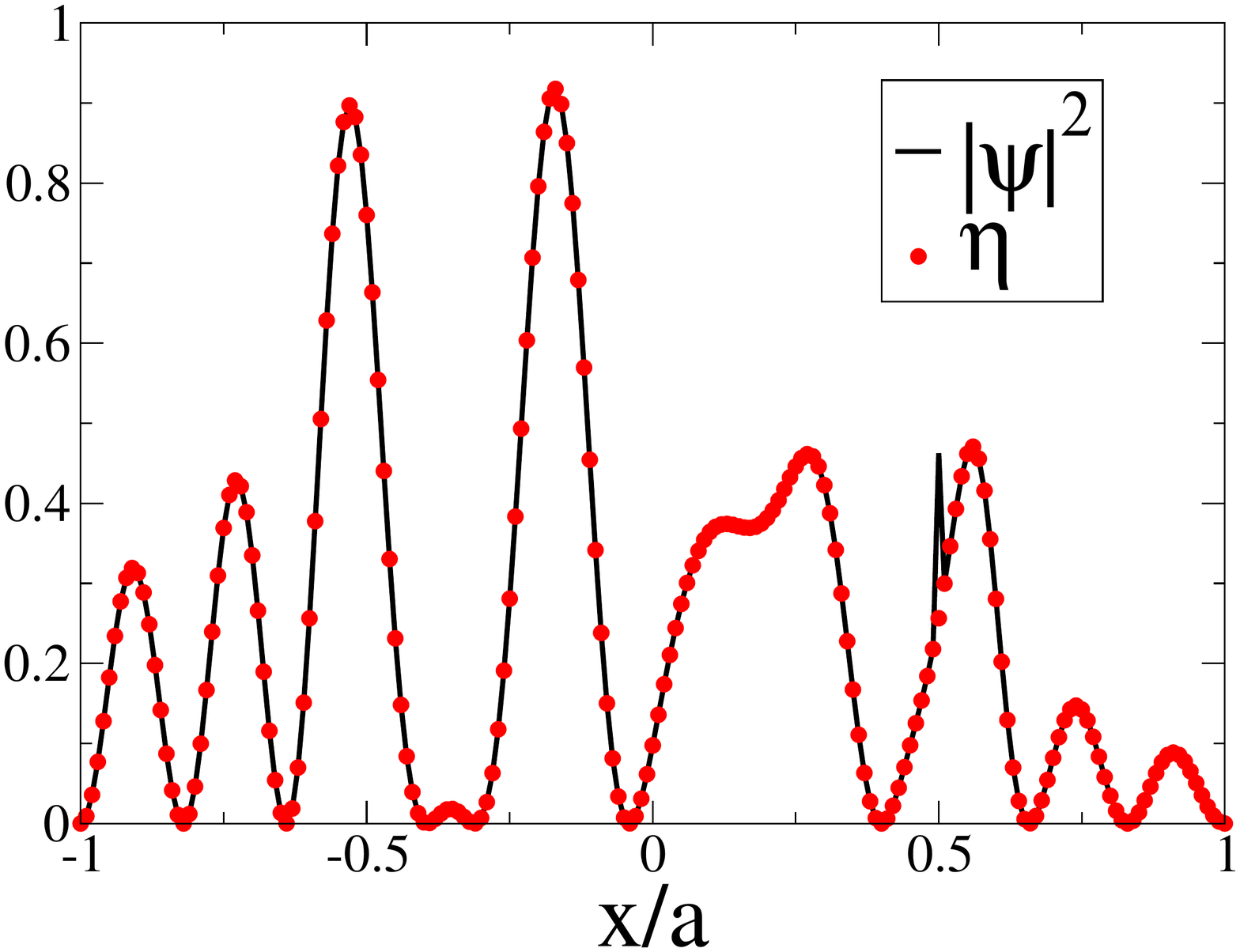}
\includegraphics[width=0.38\textwidth]{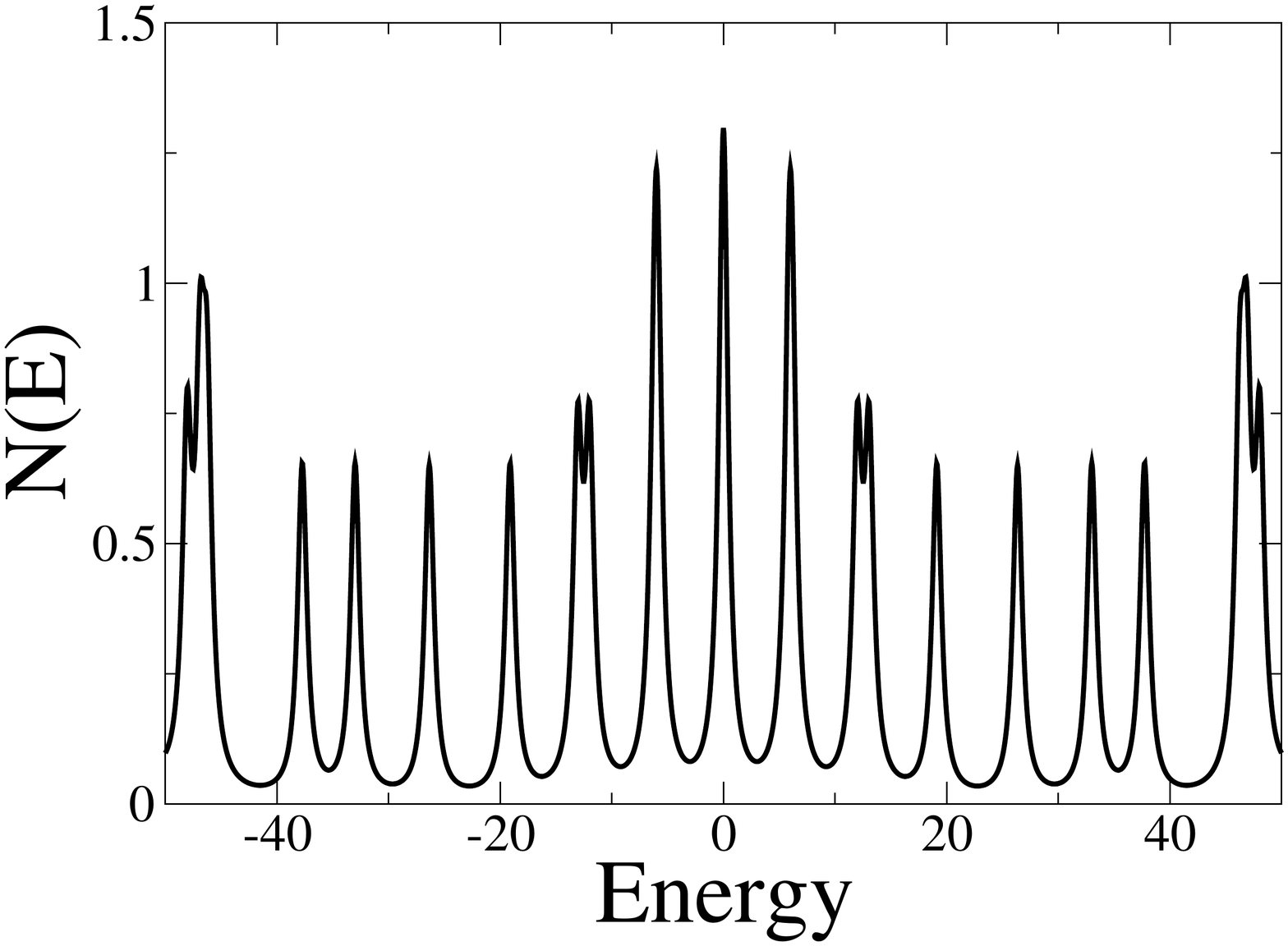}
\includegraphics[width=0.38\textwidth]{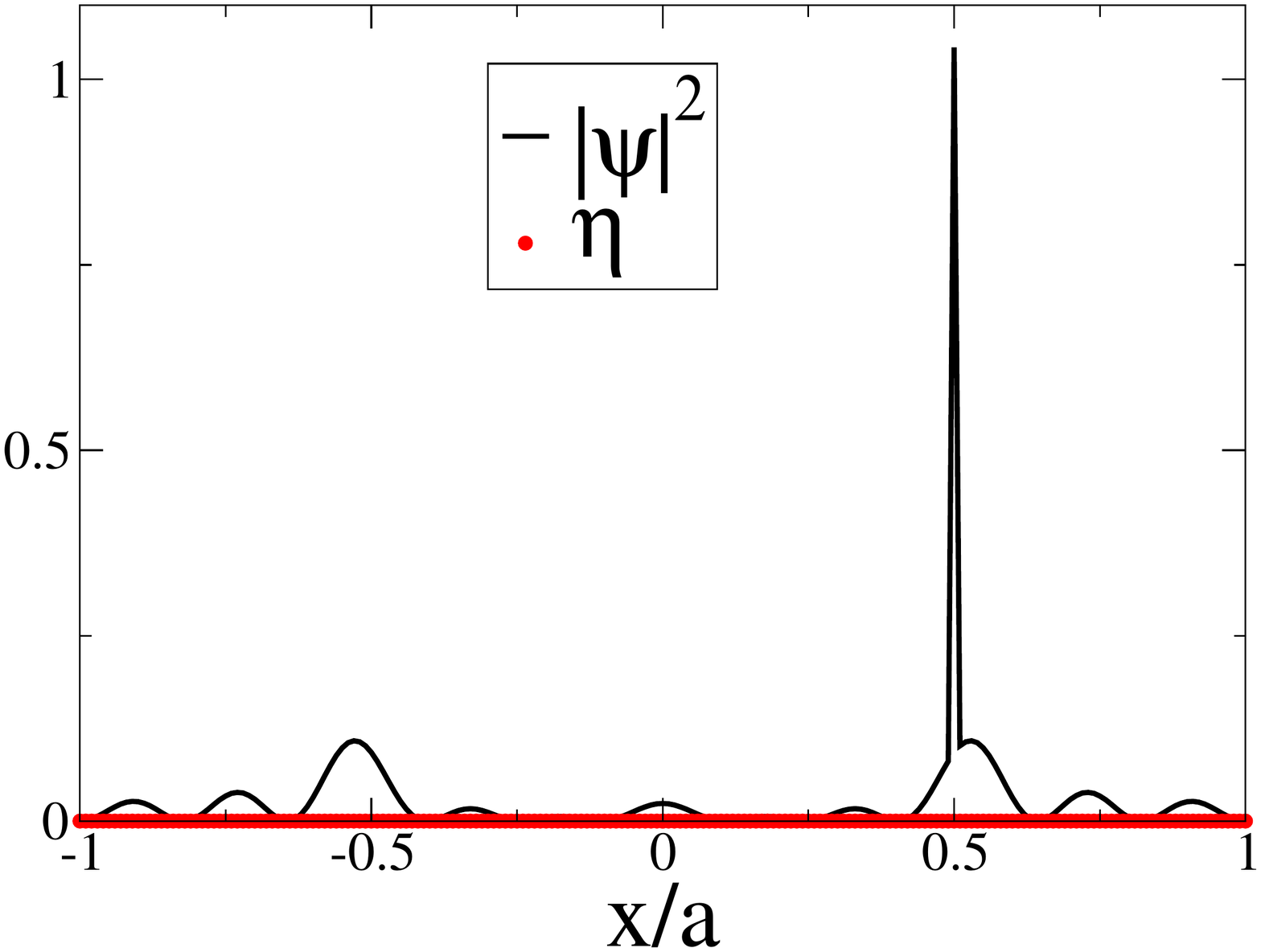}
\caption{\label{fig6}
(Color online)
Density of states (integrated over space) and self-conjugacy of the lowest energy state for the edge of 
the wire placed at the right focus for
(top panels) a fermionic mode at the edge with $t=5,t_M=0$ and (bottom panels) a Majorana mode at
the edge with $t=t_M=5$.  
}
\end{center}
\end{figure}

Further differences between the tunneling to a fermionic or to
a Majorana mode are clearly seen in the density of states obtained
integrating over space the LDOS. In the case of
tunneling to a fermionic mode there is a gap in the energy spectrum
while in the case of tunneling to a Majorana mode there is a zero energy
mode. 
The effect of the tunneling to the Majorana edge state
is better understood looking at the self-conjugacy of the wave functions indicative of a Majorana
state. Considering a tunneling with $t_M=0$, the Majorana character of the $d$ mode is destroyed while
if $t_M=t$ it is preserved, $|u_d|^2-|v_d|^2=0$. The same test can be performed calculating
at each point along $x$ the quantity $\eta(r,\varphi)=|u(r,\varphi)|^2-|v(r,\varphi)|^2$.
Considering both cases the lowest energy eigenstate has a different structure. Specifically,
when tunneling to a Majorana edge state the wave function of the lowest state is
self-conjugate accross the corral. Moreover, while in the case of the fermionic mode
tunneling the wave function has a significant amplitude accross the corral, in the
case of the tunneling to a Majorana mode the wave function is strongly peaked at the
focus location. These results are shown in Fig. \ref{fig6}.
Note that the wave functions include a contribution from the local state at the right focus.
Indeed, in the case of a Majorana mode this local mode contributes significantly with
a sharp peak, as seen in the bottom right panel.

\subsection{Differential density of states}

The effect of the hybridization to a local mode, either of fermionic or Majorana character, is better
seen calculating the differential local density of states. 
The differential local density of states is defined by the difference between the
LDOS of the uncoupled wire to the LDOS  when the wire
is hybridized with the conduction electrons.

\begin{figure}
\begin{center}
\includegraphics[width=0.42\textwidth]{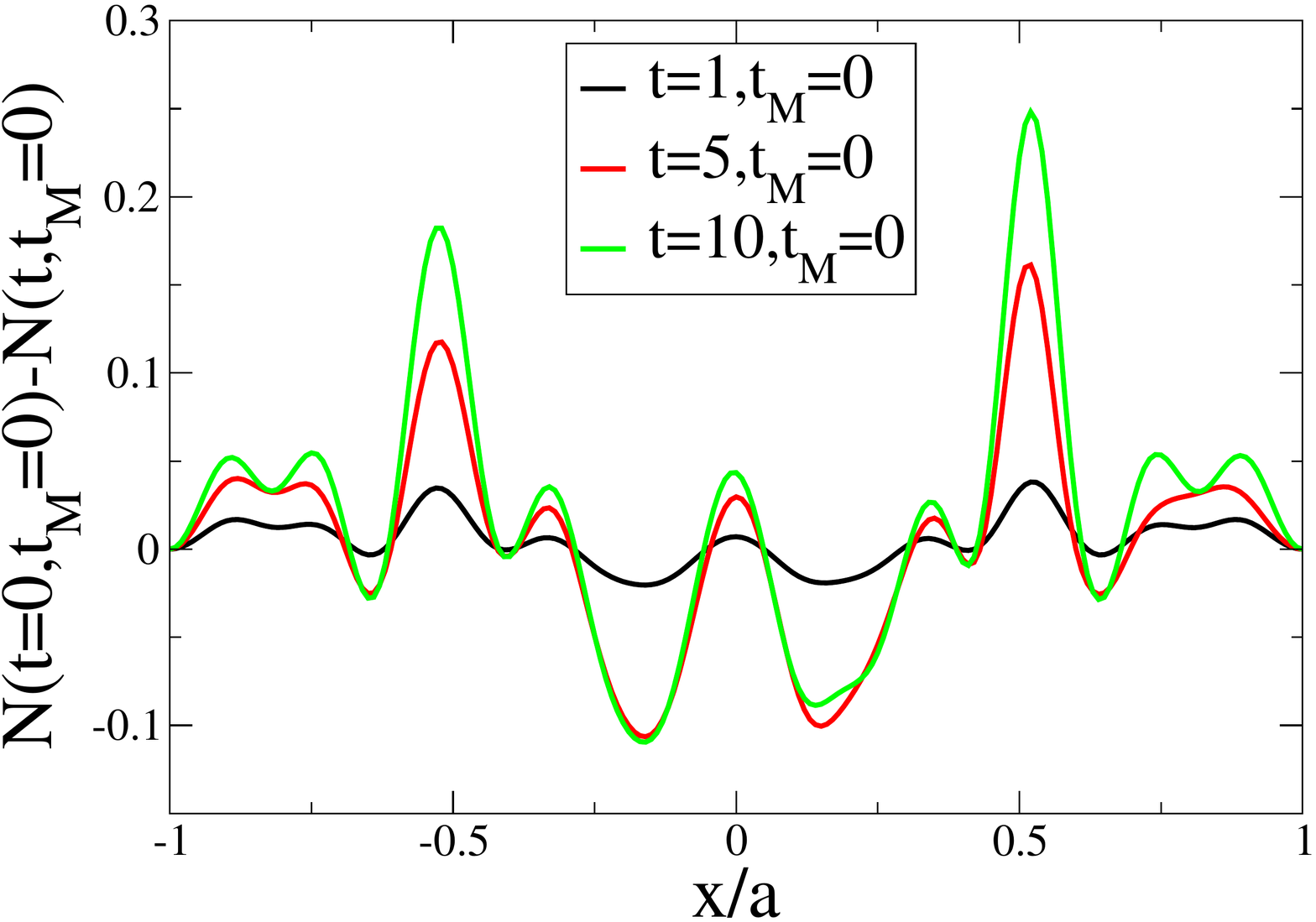}
\includegraphics[width=0.42\textwidth]{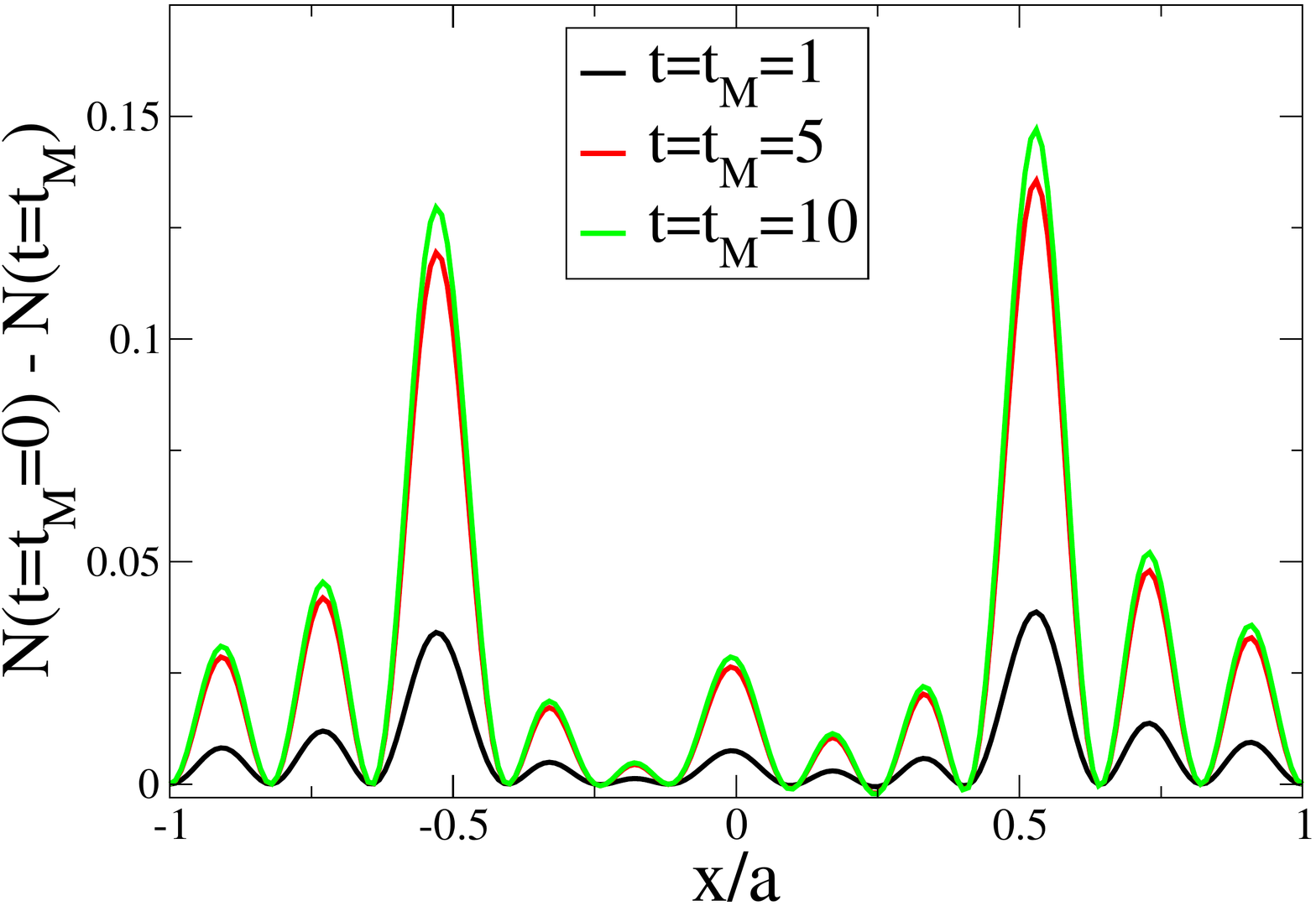}
\includegraphics[width=0.42\textwidth]{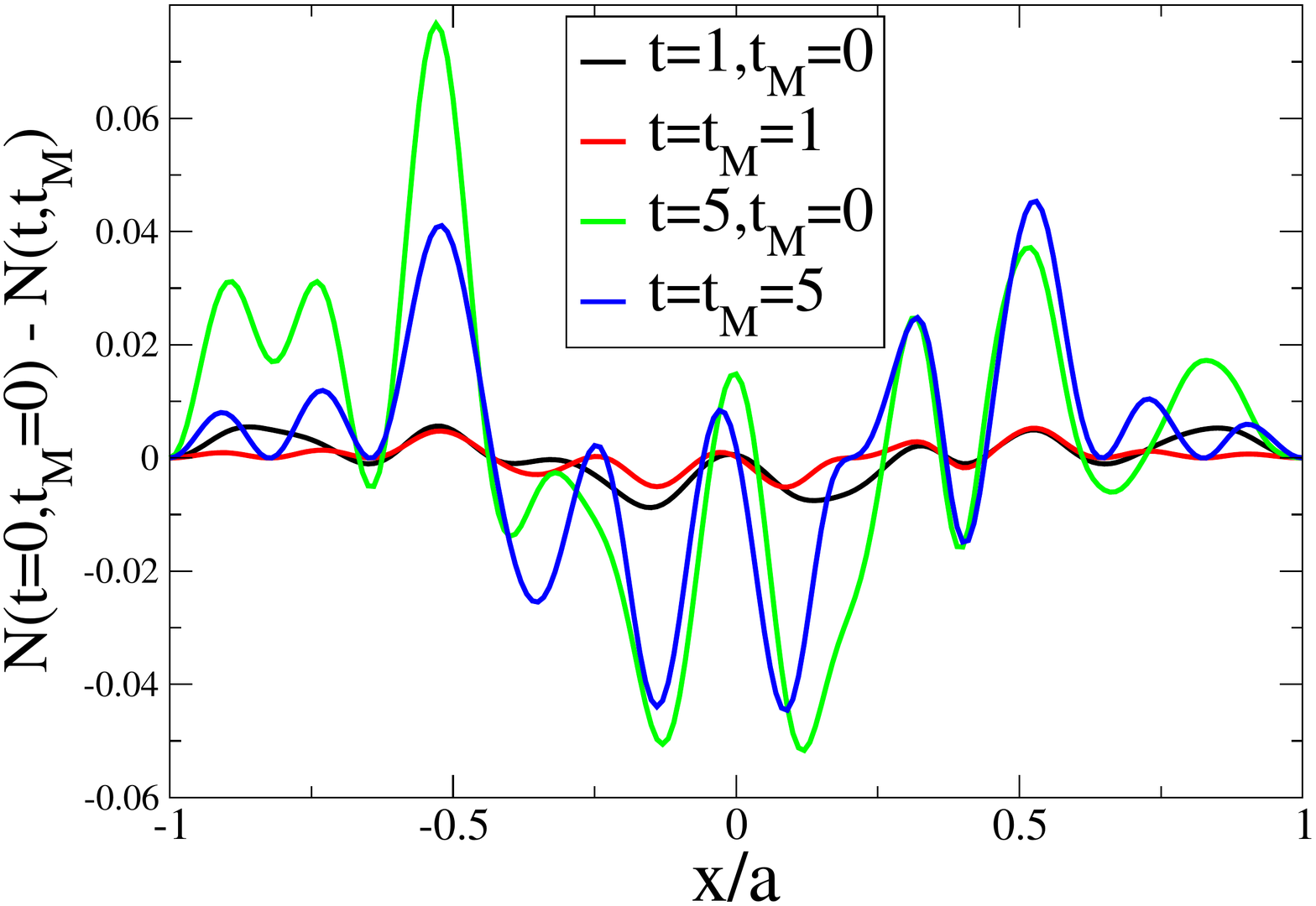}
\caption{\label{fig7}
(Color online)
Conduction electron contribution to the differential 
local density of states for a) a fermionic mode and b) a Majorana mode on focus, and c) both
modes when the wire is placed off-focus.
}
\end{center}
\end{figure}

In Fig. \ref{fig7} the differential local density of states is compared, for the cases of the edge of the wire
placed at the focus, for a fermionic and a Majorana mode. Also, the case of the local mode off focus
is considered. In the case the edge is at the focus, as the tunneling increases the LDOS 
of the conduction electrons decreases in the neighborhood of the foci, both for a fermionic and a Majorana
mode, as noticed in the previous subsection, and the differential LDOS increases significantly showing
a pronounced mirage effect. However, as referred above and made clear by the differential LDOS, the effect
in regions away from the foci is clearly different comparing a fermionic with a Majorana mode.
In the case of the Majorana mode the change in the region between the foci is small while in the case
of the fermionic mode there is a clear anti-mirage effect. 
In the case of a wire edge placed off-focus there is no clear mirage effect for either type of modes.

\section{Two wires}

The existence of the Majorana mode at the edge of the right wire 
may be detected introducing a second wire in the
system. Consider the situation where we have one wire with one edge at the right focus 
and another wire with one edge at the other focus. We may now allow the tunnelings
between the free electrons and the states at the wires to be either through a fermionic or a Majorana mode.

In Fig. \ref{fig8} a few cases are considered, since one may tune either
wire in or out of a topological phase or keeping both in trivial or
topological phases. We plot the full LDOS 
in the two-dimensional corral (including both the conduction electron and local mode contributions), 
the lowest energy eigenfunctions and their self-conjugacies. In the left and right
panels there is reflection symmetry and the groundstate is degenerate. If there are
fermionic modes at the edges of both wires ($t_R=5, t_M^R=0, t^L=5, t_M^L=0$) the
groundstate is symmetric around the center of the ellipse. However, if there are
Majorana modes, even if the tunnelings are symmetric ($t^R=t_M^R=5,t^L=t_M^L=5$),
the two wires couple and one of the wavefunctions in the lowest energy subspace of
approximately zero energy is not symmetric. However, summing the contribution from
the wave functions with zero energy symmetry is recovered.
If both at the right and left foci there is a fermionic mode, the behavior just reflects
the absence of a mirage and a high density of states in the central region of the ellipse.
Also, the lowest energy state is peaked at foci but it is not self-conjugate.
On the other hand if one of the two edge states is a Majorana, the lowest energy
state has a Majorana character (self-conjugate). The most interesting case is when both
edge states are Majoranas: there is the mirage effect mentioned above, the lowest state
wavefunction is peaked at the foci and the mode is self-conjugate.

\begin{figure*}
\begin{center}
\includegraphics[width=0.28\textwidth]{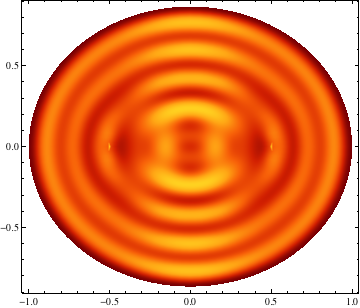}
\includegraphics[width=0.28\textwidth]{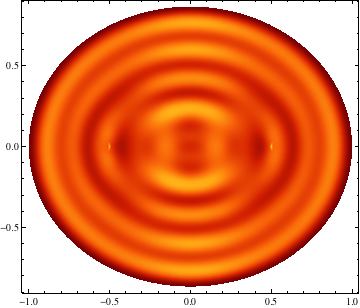}
\includegraphics[width=0.28\textwidth]{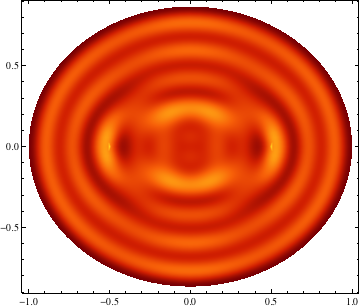}
\includegraphics[width=0.32\textwidth]{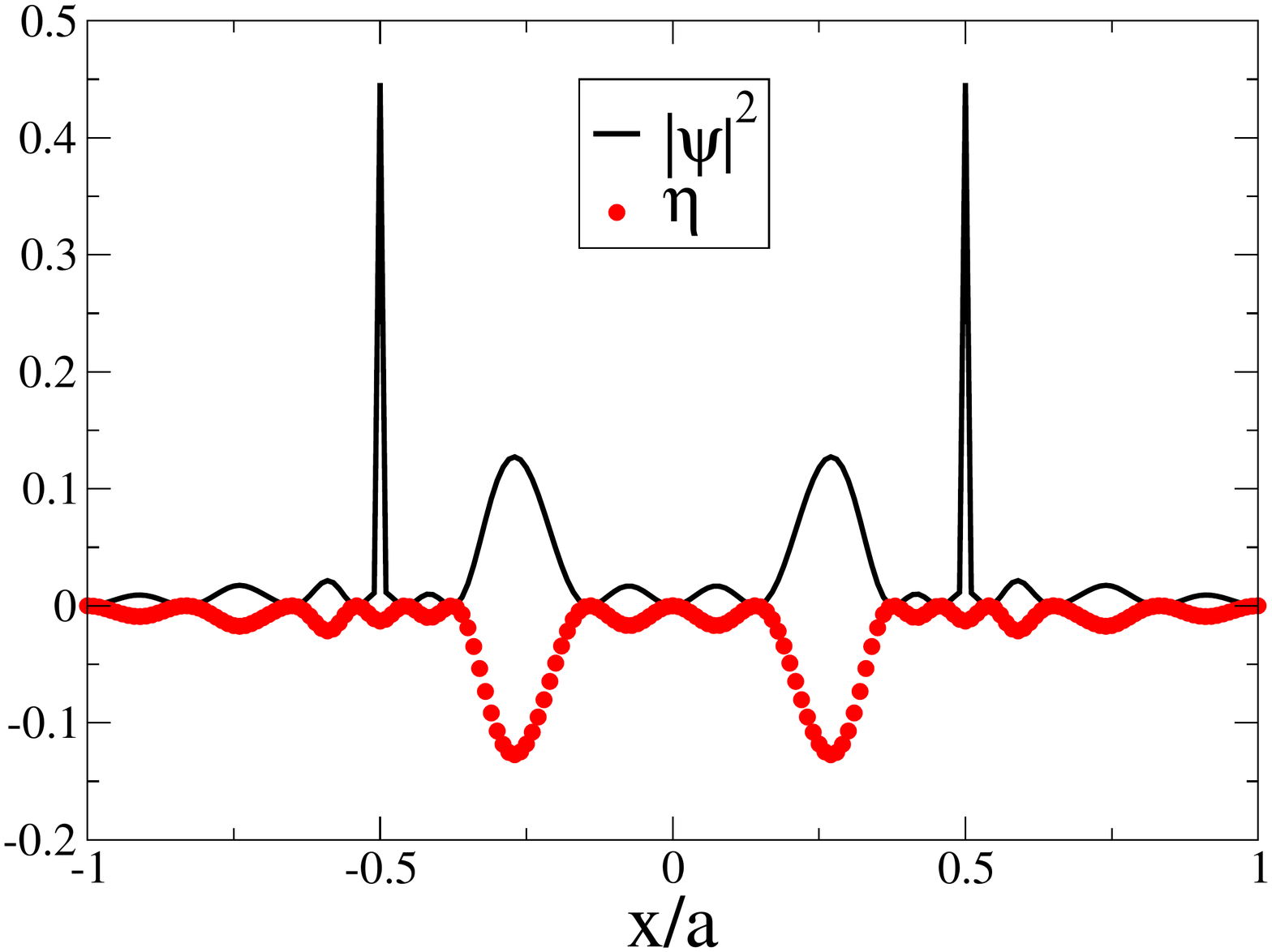}
\includegraphics[width=0.32\textwidth]{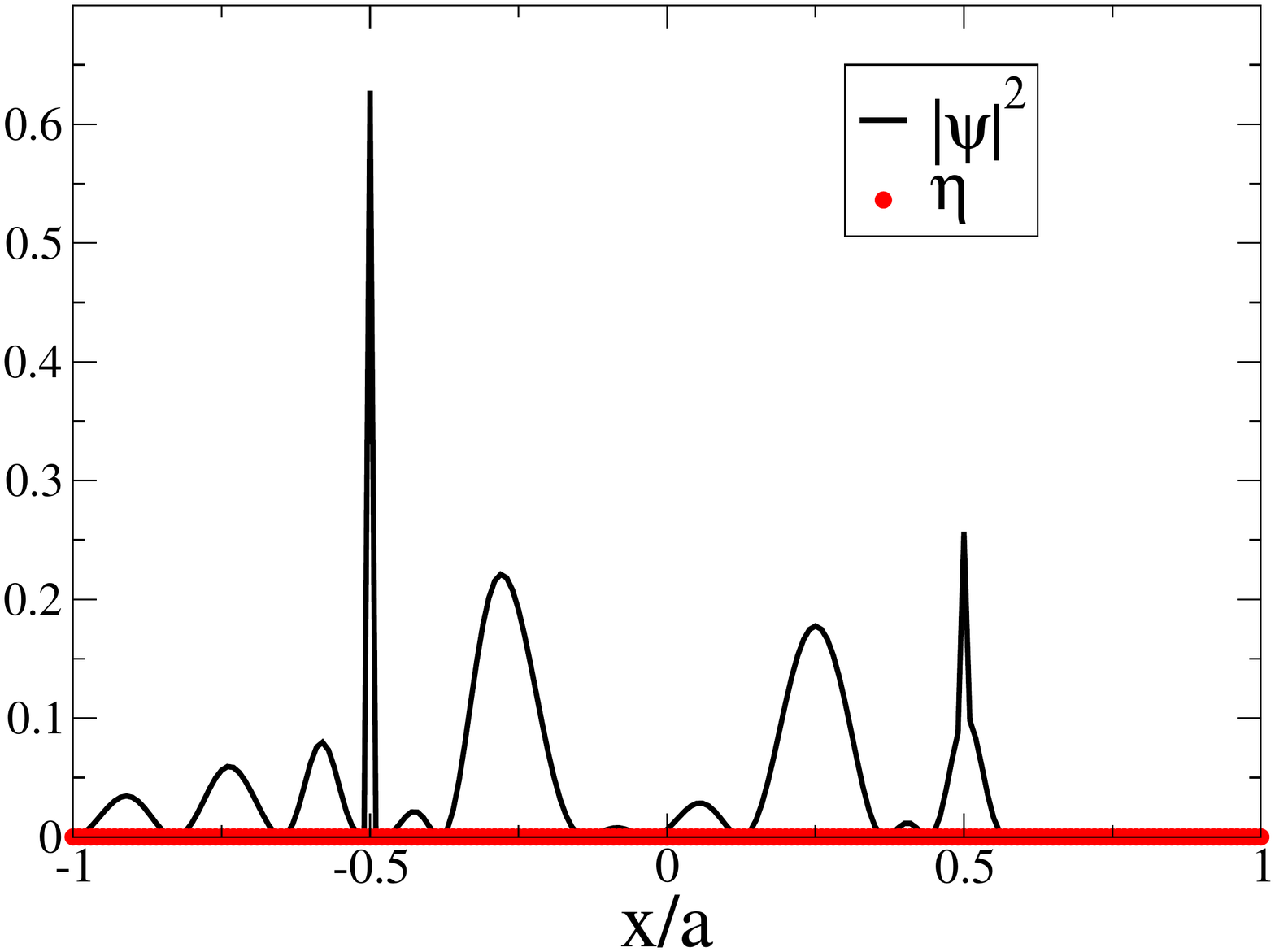}
\includegraphics[width=0.32\textwidth]{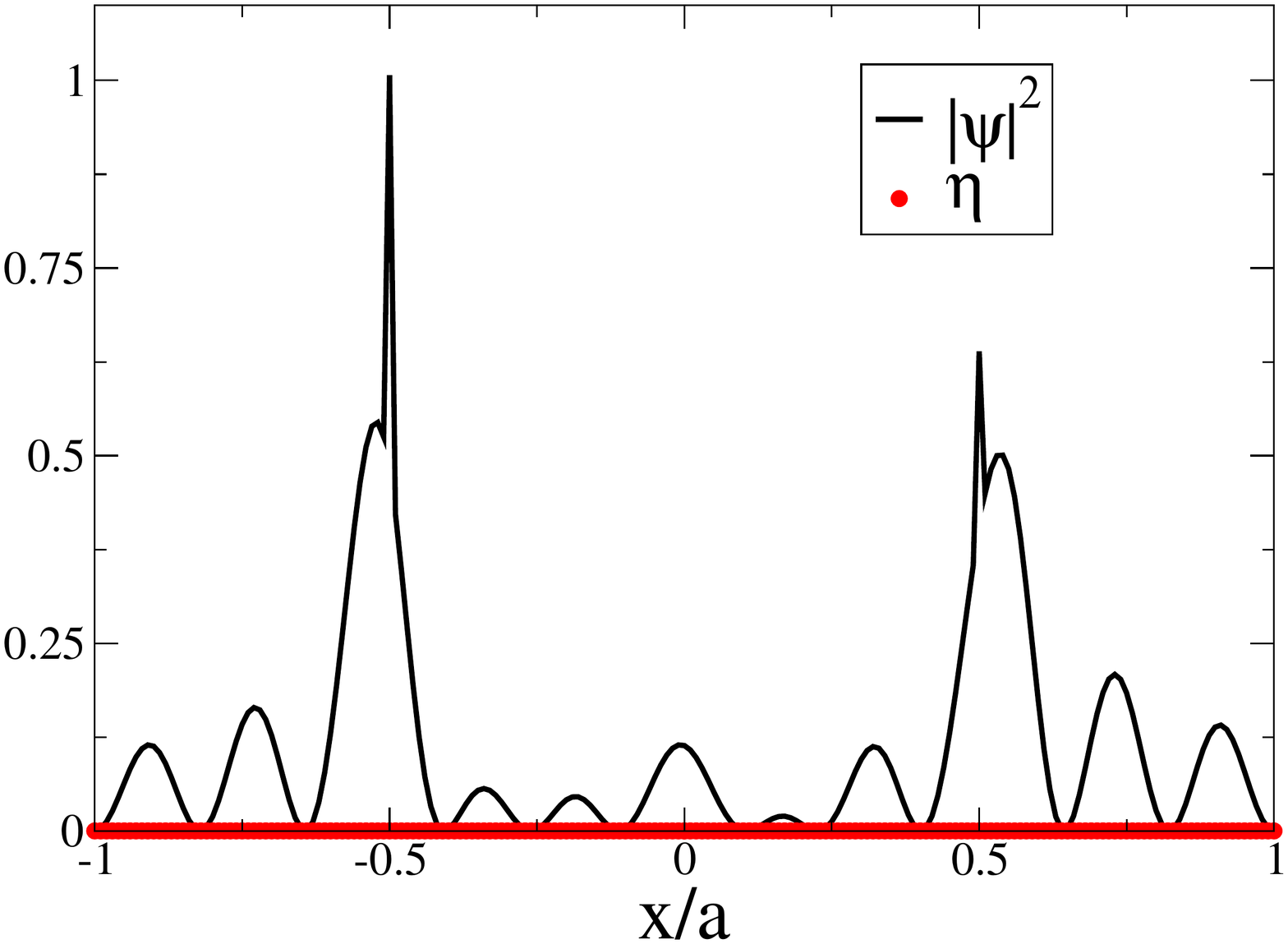}
\caption{\label{fig8}
(Color online)
Two wires at the foci: density of states, lowest energy
eigenstate amplitude and self-conjugacy for $(t^R=5,t_M^R=0;t^L=5,t_M^L=0)$,
$(t^R=t_M^R=5;t^L=5,t_M^L=0)$, and  $(t^R=t_M^R=5;t^L=t_M^L=5)$,
respectively. In the case of the Majorana modes, the wavefunction is one of a degenerate
set of solutions. 
The axis in the top panels are as in Fig. 3.
}
\end{center}
\end{figure*}

\section{Conclusions}

In summary, the tunneling between a free electron gas confined in an elliptical
corral and a topological wire with a Majorana mode placed at one focus, leads 
to a mirage effect at the other focus, given the appropriate conditions. A condition
common to all mirage problems implies that the chemical potential should be located on an
energy eigenstate with a significant amplitude at the two foci. The effect of the coupling
to the Majorana mode at the focus location, is to reduce the density of states in its vicinity,
with a peak at its exact location, and maintaining a reduced density of states
elsewhere in the corral. In the case of tunneling to a fermionic mode the mirage
effect is lost, in the sense that there is an anti-mirage effect, since the density of states
in the central region of the corral is enhanced with respect to the free system.
The results show that the choice of the chemical potential is relevant since 
placing the chemical potential on an energy level that does not have a high amplitude on both
foci, reduces the effect of the state at the wire edge. Also, placing the wire off-focus has
a reduced effect for the same reason.
The wave function of the Majorana mode, while mainly localized at the focus,
extends throughout the corral maintaining its self-conjugacy.
The difference between a coupling to a fermionic mode and a Majorana mode
is enhanced calculating the differential local density of states.
Also, the mirage effect is clearly seen in the case of placing the edge of the wire
at one focus, instead of off focus.

Superimposing two wires on the two-dimensional system and placing the edges of
each wire at the foci, leads to a mirage effect if both edge modes have a
Majorana character but an extended mode with self-conjugacy is observed if at least
one of the wires is in a topological phase with one edge Majorana.

The mirage effect may allow the teleportation of the Majorana character at arbitrary distances
and one may imagine a network of wires that may interact non-locally.
Tuning either one or both wires through a topological transition, it will be interesting
to attest the transmission of information using the non-locality of the coupling
between the Majorana modes.
Acting locally on a Majorana mode, such as in Fig. \ref{fig1} the right $\gamma_2$ Majorana,
may affect the left edge of a second wire placed on the left focus, since Majoranas occur in pairs.
The dynamics, response time and possible loss of coherence will be interesting to study.

\section{Acknowledgements}
Discussions with Pedro Ribeiro are acknowledged. 
Partial support from FCT through grant UID/CTM/04540/2013
is also acknowledged.

\begin{appendix}

\section{Hamiltonian matrix}

The Laplacian may be written as
\bea
\nabla^2 &=& \frac{\partial^2}{\partial x^2} + \frac{\partial^2}{\partial y^2}
\nonumber \\
&=& f_r(\varphi) \frac{\partial^2}{\partial r^2} +
f_{\varphi}(\varphi) \left( \frac{1}{r}  \frac{\partial}{\partial r} +
\frac{1}{r^2} \frac{\partial^2}{\partial \varphi^2} \right) + \nonumber \\
&+& f_{r \varphi}(\varphi) \left( \frac{1}{r} \frac{\partial^2}{\partial \varphi \partial r} -
\frac{1}{r^2} \frac{\partial}{\partial \varphi}\right)
\eea
where ($b\leq a$)
\bea
f_r(\varphi) &=& \frac{\cos^2 \varphi}{a^2} + \frac{\sin^2 \varphi}{b^2}
\nonumber \\
f_{\varphi}(\varphi) &=& \frac{\cos^2 \varphi}{b^2} + \frac{\sin^2 \varphi}{a^2}
\nonumber \\
f_{r \varphi}(\varphi) &=& \left( \frac{1}{b^2}-\frac{1}{a^2} \right) \sin 2\varphi
\eea
We solve Laplace's equation using the basis of the circular problem.
We consider the basis of states 
\be
\langle \vec{r}|kn\rangle = R_{kn} J_k(\gamma_{kn} r) e^{ik\varphi}.
\ee
The function $J_k$ is the cylindrical Bessel function of order $k$, $R_{kn}$ ensure
that the basis is orthonormalized and $\gamma_{kn}$ is the $n$ zero of the Bessel function
of order $k$. Imposing that
\bea
\int_0^1 r dr \int_0^{2\pi} d\varphi R_{k^{\prime} n^{\prime}} J_{k^{\prime}}(\gamma_{k^{\prime} n^{\prime}} r)
e^{-i k^{\prime} \varphi} 
R_{kn} J_k(\gamma_{kn} r) e^{i k\varphi} 
\nonumber \\
= \delta_{k,k^{\prime}} \delta_{n,n^{\prime}} \nonumber
\eea
we get that
$R_{kn}= 1/(\sqrt{\pi} J_{k+1}(\gamma_{kn}))$.

In this basis the Hamiltonian of free electrons has the following matrix
elements
\bea
&& \left( \frac{\hbar^2}{2ma^2\sigma^2} \right)^{-1} \langle k^{\prime} n^{\prime}| H_0 |kn\rangle =
\nonumber \\
&=& \delta_{kk^{\prime}} \delta_{nn^{\prime}} \frac{1}{2} (\sigma^2+1) \gamma_{k^{\prime} n^{\prime}}^2
\nonumber \\
&+& \delta_{k^{\prime},k+2} \left( \frac{\pi}{2} (\sigma^2-1) \gamma_{k^{\prime}-2,n}^2
\Gamma_{k^{\prime}n^{\prime};k^{\prime}-2,n} \right. \nonumber \\
&-& \left.
\pi (\sigma^2-1) \gamma_{k^{\prime}-2,n} (k^{\prime}-1) \Lambda_{k^{\prime}n^{\prime};k^{\prime}-2,n}^+
\right)
\nonumber \\
&+& \delta_{k^{\prime},k-2} \left( \frac{\pi}{2} (\sigma^2-1) \gamma_{k^{\prime}+2,n}^2
\Gamma_{k^{\prime}n^{\prime};k^{\prime}+2,n} \right. \nonumber \\
&-& \left.
\pi (\sigma^2-1) \gamma_{k^{\prime}+2,n} (k^{\prime}+1) \Lambda_{k^{\prime}n^{\prime};k^{\prime}+2,n}^-
\right)
\eea
The following integrals are defined
\bea
\Gamma_{k^{\prime}n^{\prime};k^{\prime}-2,n} &=& \int_0^1 r dr 
\frac{J_{k^{\prime}}(\gamma_{k^{\prime}n^{\prime}}r) }{\sqrt{\pi} J_{k^{\prime}+1}(\gamma_{k^{\prime}n^{\prime}})}
\frac{J_{k^{\prime}-2}(\gamma_{k^{\prime}-2,n}r) }{\sqrt{\pi} J_{k^{\prime}-1}(\gamma_{k^{\prime}-2,n})}
\nonumber \\
\Gamma_{k^{\prime}n^{\prime};k^{\prime}+2,n} &=& \int_0^1 r dr 
\frac{J_{k^{\prime}}(\gamma_{k^{\prime}n^{\prime}}r) }{\sqrt{\pi} J_{k^{\prime}+1}(\gamma_{k^{\prime}n^{\prime}})}
\frac{J_{k^{\prime}+2}(\gamma_{k^{\prime}+2,n}r) }{\sqrt{\pi} J_{k^{\prime}+3}(\gamma_{k^{\prime}+2,n})}
\nonumber \\
\Lambda_{k^{\prime}n^{\prime};k^{\prime}-2,n}^+ &=& \int_0^1 r dr 
\frac{J_{k^{\prime}}(\gamma_{k^{\prime}n^{\prime}}r) }{\sqrt{\pi} J_{k^{\prime}+1}(\gamma_{k^{\prime}n^{\prime}})}
\frac{J_{k^{\prime}-1}(\gamma_{k^{\prime}-2,n}r) }{\sqrt{\pi} J_{k^{\prime}-1}(\gamma_{k^{\prime}-2,n})}
\nonumber \\
\Lambda_{k^{\prime}n^{\prime};k^{\prime}+2,n}^- &=& \int_0^1 r dr 
\frac{J_{k^{\prime}}(\gamma_{k^{\prime}n^{\prime}}r) }{\sqrt{\pi} J_{k^{\prime}+1}(\gamma_{k^{\prime}n^{\prime}})}
\frac{J_{k^{\prime}+1}(\gamma_{k^{\prime}+2,n}r) }{\sqrt{\pi} J_{k^{\prime}+3}(\gamma_{k^{\prime}+2,n})}
\nonumber \\
& & 
\eea 
For $k^{\prime}=0$ we get 
\bea
\Gamma_{0 n^{\prime};-2,n} &=& -\int_0^1 r dr 
\frac{J_{0}(\gamma_{0 n^{\prime}}r) }{\sqrt{\pi} J_{1}(\gamma_{0 n^{\prime}})}
\frac{J_{2}(\gamma_{2,n}r) }{\sqrt{\pi} J_{1}(\gamma_{2,n})}
\nonumber \\
\Lambda_{0 n^{\prime};-2,n}^+ &=& \int_0^1 r dr 
\frac{J_{0}(\gamma_{0 n^{\prime}}r) }{\sqrt{\pi} J_{1}(\gamma_{0 n^{\prime}})}
\frac{J_{1}(\gamma_{2,n}r) }{\sqrt{\pi} J_{1}(\gamma_{2,n})}
\nonumber \\
& & 
\eea
and for $k^{\prime}=1$ we get
\bea
\Gamma_{1 n^{\prime};-1,n} &=& -\int_0^1 r dr 
\frac{J_{1}(\gamma_{1 n^{\prime}}r) }{\sqrt{\pi} J_{2}(\gamma_{1 n^{\prime}})}
\frac{J_{1}(\gamma_{1,n}r) }{\sqrt{\pi} J_{0}(\gamma_{1,n})}
\nonumber \\
\Lambda_{1 n^{\prime};-1,n}^+ &=& \int_0^1 r dr 
\frac{J_{1}(\gamma_{1 n^{\prime}}r) }{\sqrt{\pi} J_{2}(\gamma_{1 n^{\prime}})}
\frac{J_{0}(\gamma_{1,n}r) }{\sqrt{\pi} J_{0}(\gamma_{1,n})}
\nonumber \\
& & 
\eea
The various integrals have been calculated using Mathematica.
The infinite set of terms has been truncated and the resulting finite
matrix is diagonalized.

\end{appendix}

\nolinenumbers


\begin{thebibliography}{99}
\bibitem{evalue}
The eigenstate chosen is level $34$. Around this energy the spacing to the neighboring states
is of the order of $6-7$ in the units of $\hbar^2/\left(2ma^2\sigma^2 \right)=1$.
\end{thebibliography}
\end{document}